\documentclass[aps,prb,twocolumn,amsmath,amssymb,superscriptaddress]{revtex4-1}

\usepackage{xspace}
\usepackage{graphicx}
\usepackage{dcolumn}
\usepackage{bm}
\usepackage{color}

\newcommand{\cmi}{$\mathrm{cm}^{-1}$\xspace}
\newcommand{\BIO}{$\mathrm{Ba}_{2}\mathrm{In}_{2}\mathrm{O}_{5}$\xspace}
\newcommand{\BIOH}{$\mathrm{Ba}\mathrm{In}\mathrm{O}_{3}\mathrm{H}$\xspace}

\begin{document}

\title{Short-Range Structure of the Brownmillerite-Type Oxide \boldmath{\BIO} and its Hydrated Proton-Conducting Form \boldmath{\BIOH}}

\author{Johan Bielecki}
\email{johanb@xray.bmc.uu.se}
\affiliation{Department of Applied Physics, Chalmers University of Technology, SE-41296 G{\"o}teborg, Sweden}
\affiliation{Department of Cell and Molecular Biology, Uppsala University, Box 596, SE-75124 Uppsala, Sweden}
\author{Stewart F. Parker}
\affiliation{ISIS Facility, STFC Rutherford Appleton Laboratory, Chilton, Didcot, Oxon OX11 0QX UK}
\author{Dharshani Ekanayake}
\affiliation{Department of Applied Physics, Chalmers University of Technology, SE-41296 G{\"o}teborg, Sweden}
\author{Seikh Mohammad Habibur Rahman}
\affiliation{Department of Chemical and Biological Engineering, Chalmers University of Technology, SE-41296 G{\"o}teborg, Sweden}
\author{Lars B\"{o}rjesson}
\affiliation{Department of Applied Physics, Chalmers University of Technology, SE-41296 G{\"o}teborg, Sweden}
\author{Maths Karlsson}
\email{maths.karlsson@chalmers.se}
\affiliation{Department of Applied Physics, Chalmers University of Technology, SE-41296 G{\"o}teborg, Sweden}
\date{\today}


\begin{abstract}
The vibrational spectra and short-range structure of the brownmillerite-type oxide \BIO and its hydrated form \BIOH, are investigated by means of Raman, infrared, and inelastic neutron scattering spectroscopies together with density functional theory calculations. 
For \BIO, which may be described as an oxygen deficient perovskite structure with alternating layers of InO$_{6}$ octahedra and InO$_{4}$ tetrahedra, the results affirm a short-range structure of $Icmm$ symmetry, which is characterized by random orientation of successive layers of InO$_{4}$ tetrahedra. 
For the hydrated, proton conducting, form, \BIOH, the results suggest that the short-range structure is more complicated than the $P4/mbm$ symmetry that has been proposed previously on the basis of neutron diffraction, but rather suggest a proton configuration close to the lowest energy structure predicted by Martinez \emph{et al.} [J.-R. Martinez, C. E. Moen, S. Stoelen, N. L. Allan, J. of Solid State Chem. \textbf{180}, 3388, (2007)].
An intense Raman active vibration at 150 \cmi is identified as a unique fingerprint of this proton configuration.

\end{abstract}

\maketitle

\section{Introduction}
\vspace{-1mm}
Acceptor doped proton conducting perovskite type oxides, of the form $AB$O$_{3}$, are currently receiving considerable attention due to their potential to be used as proton conducting electrolytes in various electrochemical devices, such as hydrogen sensors, electrochromic displays and next-generation environmentally friendly fuel cells.\cite{KRE96,KRE03,Malavasi,Mathsdalton}
The acceptor doping, such as In$^{3+}$ substituted for Zr$^{4+}$ in BaZrO$_{3}$, creates an oxygen deficient structure (BaZr$_{1-x}$In$_{x}$O$_{3-x/2}$), that can be hydrated by treatment in a humid atmosphere at elevated temperatures.\cite{KRE03} 
During this process, water molecules in the gaseous phase dissociate into -OH$^{-}$ groups, which fill the oxygen vacancies, and protons (H$^{+}$), which bind to lattice oxygens, thus forming ideally compounds of the form BaZr$_{1-x}$In$_{x}$O$_{3}$H$_{x}$.\cite{KRE03}
The protons are not stuck to any particular oxygens, but can at elevated temperatures jump from one oxygen to another, which results in these materials' high proton conductivities.\cite{KRE03} 
However, details about the mechanistic aspects of the proton conduction and how it depends on the short-range structure of the material are not fully understood. 
Such fundamental understanding is key for the development of new, more highly conducting, materials, which is crucial for the development of most of the technological devices mentioned above.

In previous papers, some of us investigated the short-range (local) structure of the proton conducting perovskite type oxide system BaZr$_{1-x}$In$_{x}$O$_{3-x/2}$ ($x$ = 0--0.75), using Raman and infrared (IR) spectroscopies,\cite{KAR08_Shortrange} and inelastic neutron scattering (INS).\cite{KAR08_Tosca}
In the present work, we expand the dopant modification and focus on the Raman and IR spectra of the brownmillerite structured end member \BIO, and add density functional theory (DFT) calculations to assist in the interpretation of experimental data.

\begin{figure*}[t]
\includegraphics[width=0.88\linewidth]{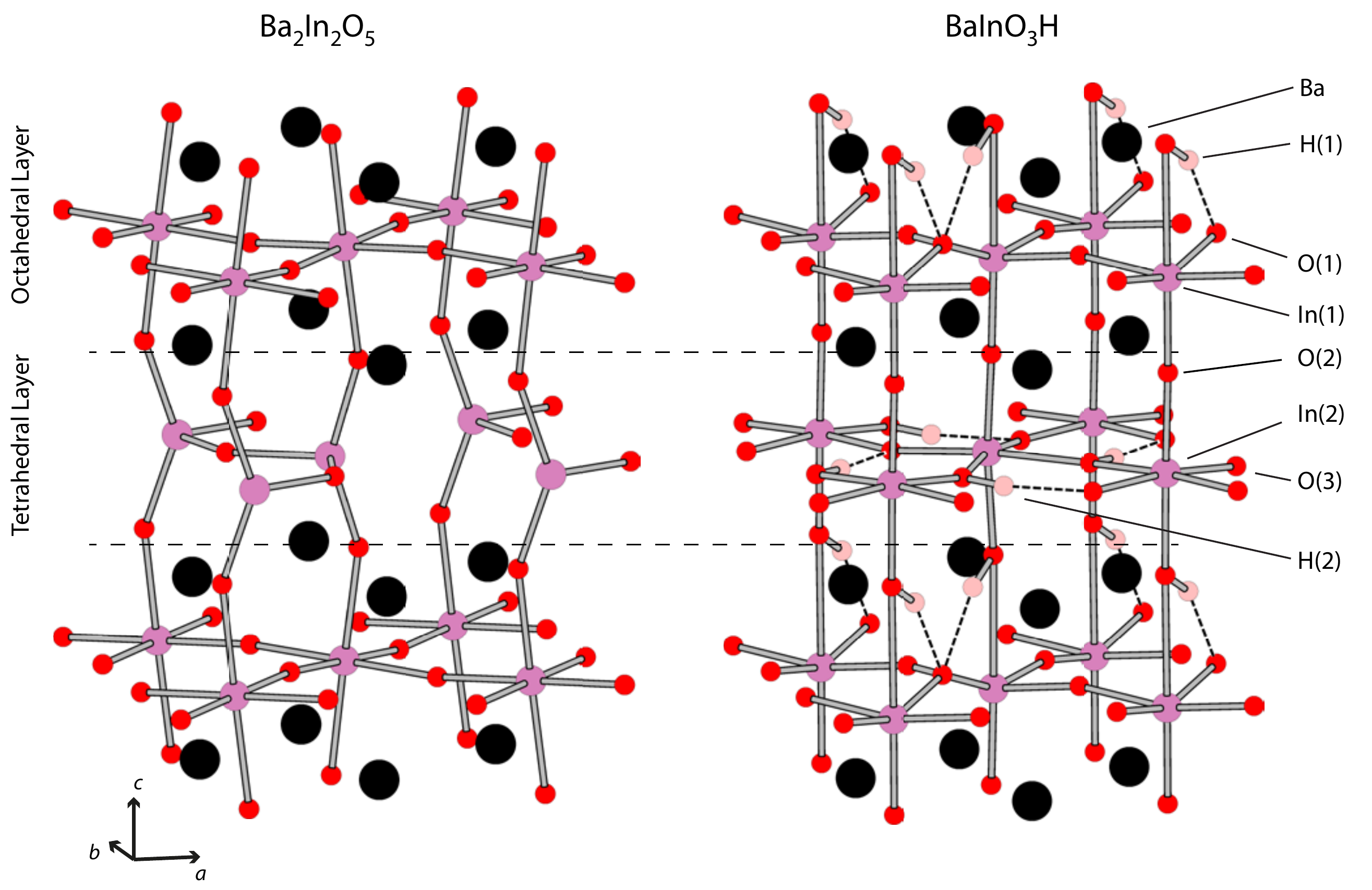}
\caption{Schematic illustration of the $Ibm2$ structure of \BIO (left) and of the $Martinez1$ structure of \BIOH (right), as used in the DFT calculations, together with atomic labels. Ba is depicted by black, O by red, In by pink, and H by white spheres, whereas hydrogen bonds are illustrated with dotted lines between O and H. 
We note that even though \BIOH is solely built up of InO$_{6}$ octahedra, we will keep the  ``tetrahedral''  and ``octahedral'' nomenclature to distinguish these two layers from each other.  
}
\label{fig:Structure}
\end{figure*}

The brownmillerite structure of \BIO may be described as an oxygen deficient variant of the perovskite structure, with alternating layers of InO$_{6}$ octahedra and InO$_{4}$ tetrahedra.
As shown on the left-hand side of Fig. \ref{fig:Structure}, the tetrahedral layers contain the In(2) and O(3) atomic positions and the octahedral layers contain the In(1) and O(1), with the two types of layers bridged by apical oxygens, denoted O(2). 
Depending on the ``handedness'', \textit{i.e.} the relative orientation of successive layers of oxygen tetrahedra, denoted either $R$ or $L$ in Fig.~1, the brownmillerite structure may be characterized by a number of different space groups, with the most common being $Ibm2$ with the same, $Pnma$ with alternating, and $Icmm$ with random handedness,\cite{Ramezanipour} as shown schematically in Fig. \ref{fig:TetHand}. 
\BIO has been reported to belong to space group $Icmm$,\cite{Berastegui,Speakman,Jiang13} however it is not clear to which extent this is due to uncoupled $Ibm2$ domains, extended defects, or actual random handedness on the local scale.\cite{Berastegui}

Upon hydration, a structural phase transition occurs towards the layered, proton conducting, \BIOH material, which is characterized by successive layers of InO$_{6}$ octahedra.\cite{KAR08_Shortrange}
Results obtained from Rietveld refinement of neutron diffraction data of \BIOH suggest its space group to be of $P4/mbm$ symmetry,\cite{Jayaraman} however, the presence of potential short-range structural distortions may lead to a short-range structure different from that average structure which has been identified by diffraction.\cite{Martinez} 
For this reason, we here investigate the short-range structures of \BIO and \BIOH and how these relate to subtle variations in the arrangement of InO$_{6}$ octahedra and InO$_{4}$ tetrahedra and protons. 
Using periodic DFT calculations we examine several possible structures and the calculated spectra are then compared with those observed with INS, and Raman and IR spectroscopy. 
This is a stringent test of the examined models and provides insight into the local structures present and their vibrational fingerprints. Moreover, we relate our results to the mechanism of proton conductivity in \BIOH, and our calculations enable us to provide a complete phonon assignment of all Raman and IR active modes for both materials.

\section{Experimental}

\subsection{Sample preparation}

Ba$_2$In$_2$O$_5$ powder was prepared by solid state sintering by mixing stoichiometric amounts of BaCO$_3$ and In$_2$O$_3$, with the sintering process divided into three treatments: 1000~$^\circ$C for 8 h, 1200 $^\circ$C for 72 h, and at 1325~$^\circ$C for 48 h, with intermediate cooling, grinding and compacting of pellets between each heat treatment.
The as-sintered Ba$_2$In$_2$O$_5$ powder was annealed in vacuum at high temperature ($\sim$600~$^\circ$C) in order to remove any protons that the sample may have taken up during its exposure to ambient conditions; this sample is referred to as dehydrated.
The hydrated sample, \BIOH, was prepared by annealing a portion of the dehydrated sample at $\sim$300~$^{\circ}$C under a flow of N$_{2}$ saturated with water vapor for a period over a few days. 
Thermal gravimetric analysis [Fig.~S1(a)] indicated that this sample was found to be fully hydrated, whereas X-ray diffraction patterns of both samples  were in agreement with the structures as reported earlier [Fig.~S1(b)].

\begin{figure}
\includegraphics[width=1\columnwidth]{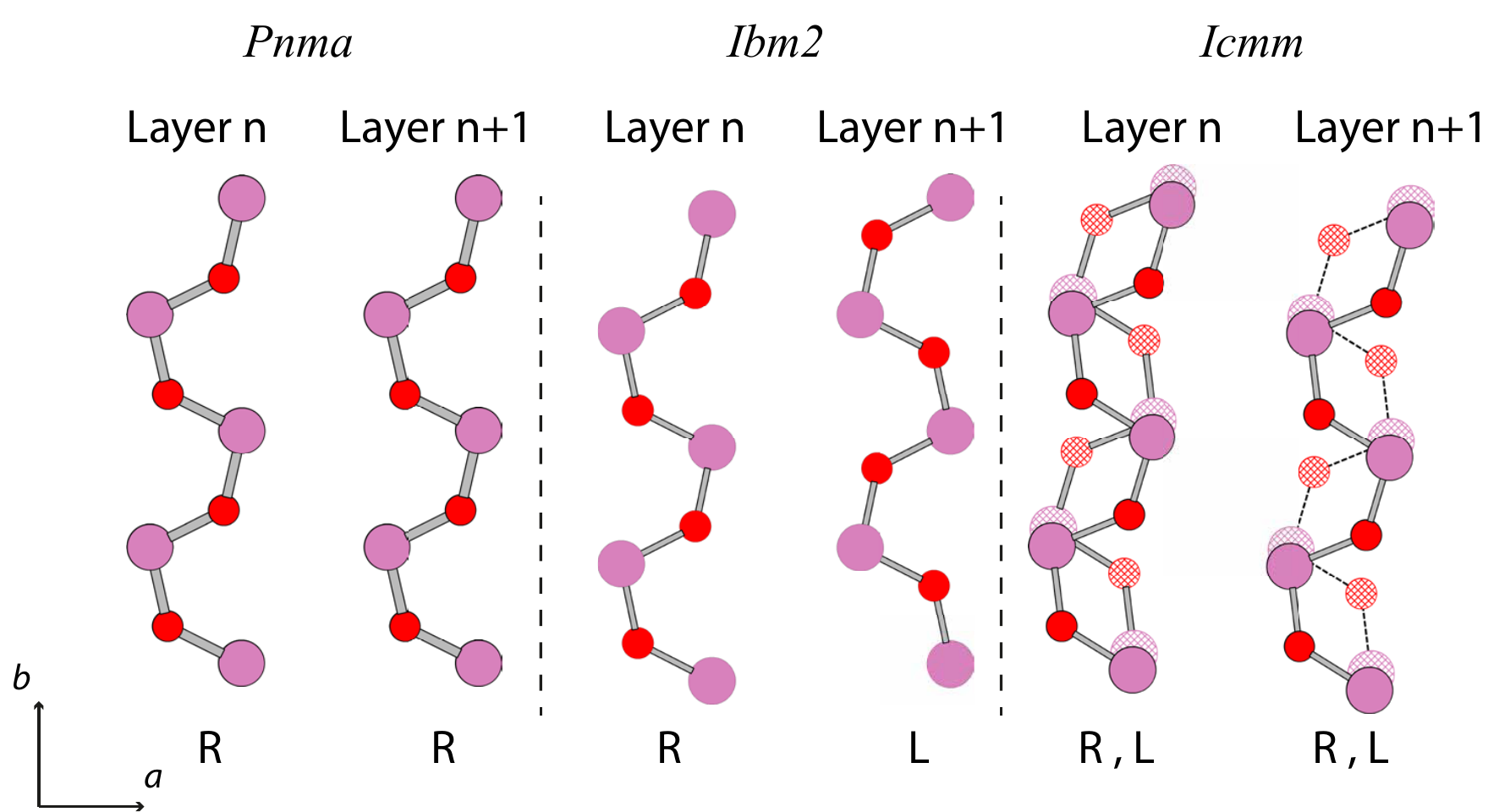}
\caption{Schematic illustration of the orientations in successive tetrahedral layers along the $c$-axis for different space groups of the brownmillerite structure.
In $Ibm2$ the same orientation is repeated indefinitely. 
In $Pnma$ the orientation alternates between adjacent layers. 
In $Icmm$ there are no orientational correlations between adjacent layers. }
\label{fig:TetHand}
\end{figure}

\subsection{Raman spectroscopy}

Raman spectra over the range 50--3800 \cmi were collected at 300 K with a DILOR XY-800 spectrometer, equipped with a charge couple device (CCD) detector, in a double subtractive grating configuration. 
The source of excitation was the 514.5 nm line of an $\mathrm{Ar}^+$ laser, which was focussed on the sample through a long working distance x40 objective.
 The power on the sample was adjusted to 4 mW, with a laser spot size of around 1 $\mu$m in diameter. 
 A comparison of Stokes and anti-Stokes spectra showed negligible laser heating on the sample during all measurements. 
 All spectra were measured  with linearly polarized light impinging on the sample and unpolarized light collected at the CCD. 

\subsection{Infrared spectroscopy}

Infrared spectra were measured over the range 75--4000~cm$^{-1}$, using three different experimental setups.
The 75--800 \cmi range was collected in transmission mode using a Bruker IFS 66v/s spectrometer, equipped with a deuterated triglycerine sulfate (DTGS) detector and exchangeable beam splitters; a Mylar 6 beam splitter was used for the 75--450 \cmi range and a KBr beam splitter for the 450--800 \cmi range.
The powder samples were dispersed to $\sim$5 wt.\% in 0.1 g of polyethylene powder (75--450 \cmi) and 0.1 g KBr powder (450--800 \cmi), respectively, and thereafter pressed into cylindrical pellets under a load of 7 tons. 
The spectrum of a wrinkled aluminium foil was used as reference spectrum. 
An absorbance-like spectrum was derived by taking the logarithm of the ratio between the reference spectrum and the sample spectrum. 
All measurements were performed at room temperature.

\subsection{Inelastic neutron scattering spectroscopy}

The sample of \BIOH  was loaded into an aluminium sachet and the sachet into an indium wire sealed thin-walled aluminium can. This was run on MAPS \cite{MAPS} at ~10 K with an incident energy, E$_i$, of 650 meV with the Fermi chopper at 600 and 500 Hz and also with E$_i$ = 300 meV and 400 Hz.  The same sample was also measured on TOSCA \cite{TOSCA}  at 10 K. For the present work, the key differences are that MAPS allows access to the O-H stretch region, enables both momentum transfer ($Q$, \AA$^{-1}$) and and energy transfer ($\omega$, \cmi) to be measured independently and is more sensitive while TOSCA has better resolution in the region below 1600 \cmi.  It was not possible to record INS spectra of \BIO as the combination of the small total cross section of the material combined with the absorption of $^{115}$In (202 barns, 95.72\% natural abundance) resulted in unacceptably poor quality spectra. 
For \BIOH the presence of hydrogen with its $\sim$100-fold larger cross section enabled good quality spectra to be measured in a reasonable time. 

\subsection{Density functional theory calculations}

Periodic DFT calculations were performed using the plane-wave pseudopotential method implemented in the CASTEP code.\cite{CASTEP1,CASTEP2} 
Exchange and correlation effects were approximated using the PBE functional.\cite{PBE} 
A norm-conserving pseudopotential with a plane-wave cut-off energy of 750 eV was used. The equilibrium structure, an essential prerequisite for lattice dynamics calculations was obtained by Broyden$-$Fletcher$-$Goldfarb$-$Shanno geometry optimization after which the residual forces were converged to zero within 0.009 eV/\AA. Phonon frequencies were obtained by diagonalisation of dynamical matrices computed using density-functional perturbation theory (DFPT).\cite{CASTEP2} 
DFPT was also used to compute the dielectric response and the Born effective charges, and from these the mode oscillator strength tensor and infrared absorptivity were calculated. The Raman activity tensors were calculated using a hybrid finite displacement/DFPT method.\cite{CASTEP3}
Raman spectra were generated from the CASTEP output using the program DOS.pl, which is part of the PHONON utility,\cite{CASTEP} assuming a temperature of 300 K and a laser excitation wavelength of 514.5 nm.

A calculation of the phonons in a solid requires fully occupied sites, thus the fractional atomic occupancies in the proposed $Icmm$ symmetry of \BIO are inappropriate for a direct DFT calculation. The calculations were instead performed in the $Pnma$ and $Ibm2$ realizations of \BIO. 
It should be noted, however, that the atomic arrangements are very similar between these three structures, differing only in the tetrahedral ordering as exemplified in Fig. \ref{fig:TetHand}, and any similarities and discrepancies between the two calculations and the experimental spectra may therefore be utilized to deduce detailed information about the short-range structure of \BIO. 

Fractional occupancies appear also for the H(1) protons in the $P4/mbm$ symmetry proposed for \BIOH, where they reside at the 16$l$ Wyckoff position with $1/8$ occupancy.
 However, in the more realistic local proton arrangements found by Martinez \emph{et al.} \cite{Martinez}, the 16$l$ protons shift to the 32$y$ position at the same time as the H(2) protons at the 2$c$ position move to the new 4$h$ position, where 4$h$ and 32$y$ represent local deviations from the average 2$c$ and 16$l$ positions. These structures can be described with full occupancies given that the overall symmetry is artificially lowered to $P1$. For this reason, calculations in the hydrated phase can only be reliably compared to the INS data, where the broken symmetry will not affect any selection rules. This is opposed to the calculated Raman and IR spectra, where care has to be taken when comparing to the experimental spectra as all vibrational modes will show as active.

\section{Results}
\label{sec:Results}

\subsection{INS spectra}
\begin{figure}[t]
\includegraphics[width=1\columnwidth]{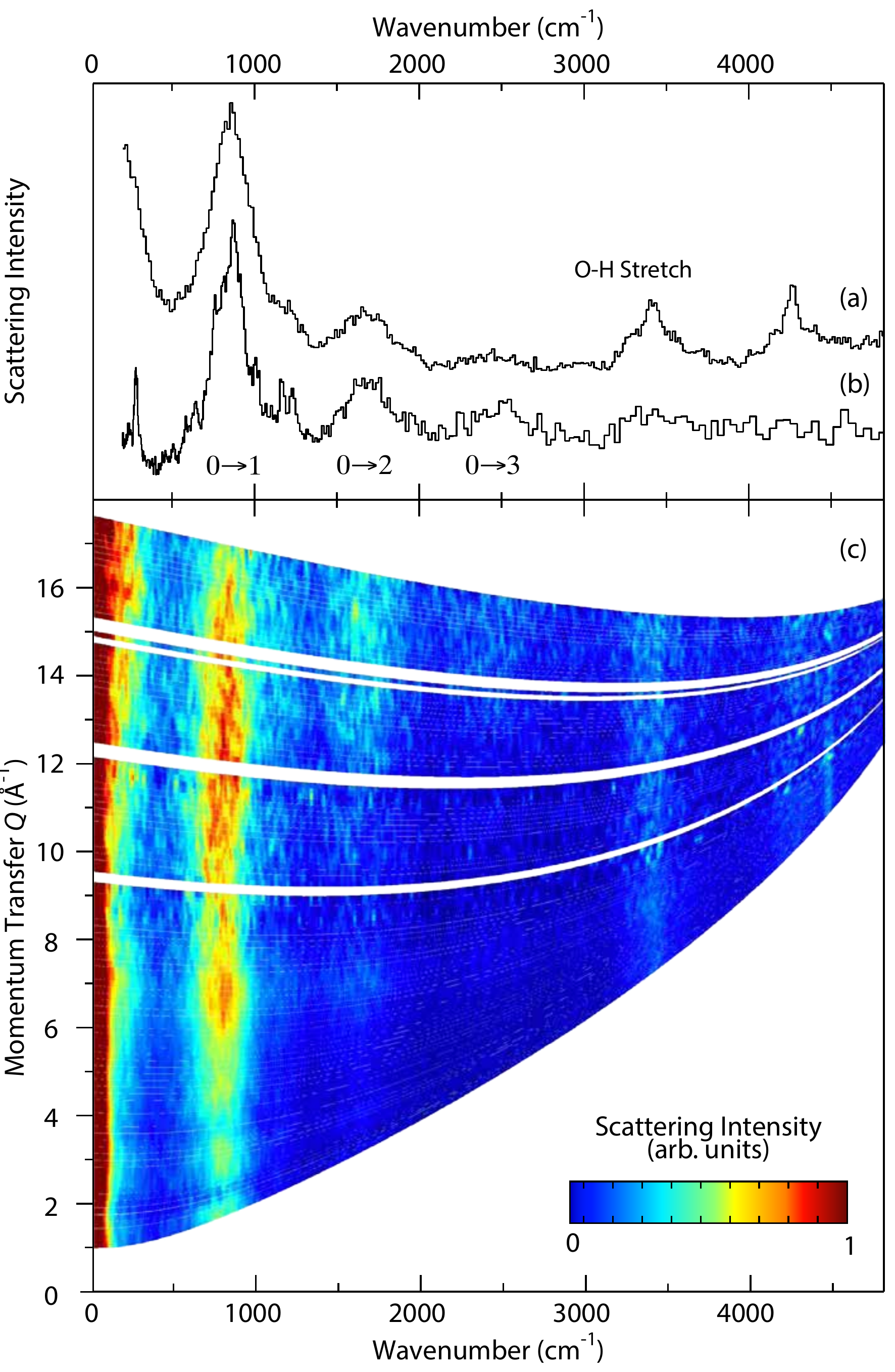}
\caption{INS spectrum at 10 K of  \BIOH recorded with (a) MAPS (E$_i$ = 650 meV) and (b) TOSCA. (c) $S(Q,\omega)$ map of  \BIOH recorded on MAPS (E$_i$ = 650 meV).}
\label{fig:INS}
\end{figure}

Figure~\ref{fig:INS} shows the INS spectrum at 10 K of \BIOH recorded with MAPS (E$_i$ = 650 meV)  and TOSCA. 
Figure~\ref{fig:INS}(a) clearly shows an O-H stretch band peaking at around 3410 \cmi. 
The O-H stretch band is broad and asymmetric, with shoulders at 3280 \cmi and 3540 \cmi. 
The intense INS feature at 860 \cmi, is assigned as the O-H wag mode fundamental (0 $\to$ 1 transition) and the weaker and broader features at 1662 and 2414 \cmi are assigned as the first (0 $\to$ 2 transition) and second overtones (0 $\to$ 3 transition) respectively. 
This assignment is in agreement with the INS spectra of the similar materials BaZr$_{1-x}$In$_{x}$O$_{3-x/2}$ ($x$ = 0.20--0.75) as reported previously,\cite{KAR08_Tosca} and is further supported by the momentum transfer, $Q$, dependence of the modes as seen in Fig.~\ref{fig:INS}(c). 
The maximum intensity of the mode occurs at larger $Q$ as the transition energy increases, exactly as expected for an overtone progression.\cite{Rb2PtH6} 
The O-H stretch band discussed above could be assigned to either hydroxyls or water (or both), however, the absence of a low $Q$ bend mode at ~1600 \cmi shows that only hydroxyls are present. 
Furthermore, the broad feature at 4270 \cmi, which is also observed in the IR spectra of  the similar materials BaZr$_{1-x}$In$_{x}$O$_{3-x/2}$ ($x$ = 0.25--0.75),\cite{KAR05} is most likely associated with a combination of O-H stretch and bend modes.
This combination is commonly seen in inorganic hydroxyl containing materials $e.g.$ LiOH$\cdot$H$_{2}$O\cite{LiOH}  and hydroxyapatite\cite{HAP} and provides further evidence for the assignment as hydroxyls. 

Figure~\ref{fig:INS}(b) shows the spectrum recorded on  TOSCA, this is similar to that obtained on MAPS but with higher spectral resolution in the low frequency region. 
The O-H wag mode at 869 \cmi is asymmetric with a pronounced shoulder at 761 \cmi and has several weaker peaks associated with it: 633, 1008, 1163 and 1227 \cmi. Such spread in the O-H wag frequency points towards several different O-H distances and at least two unique proton positions inside the crystal structure.

\subsection{IR spectra}
 
\begin{figure}
\includegraphics[width=1\columnwidth]{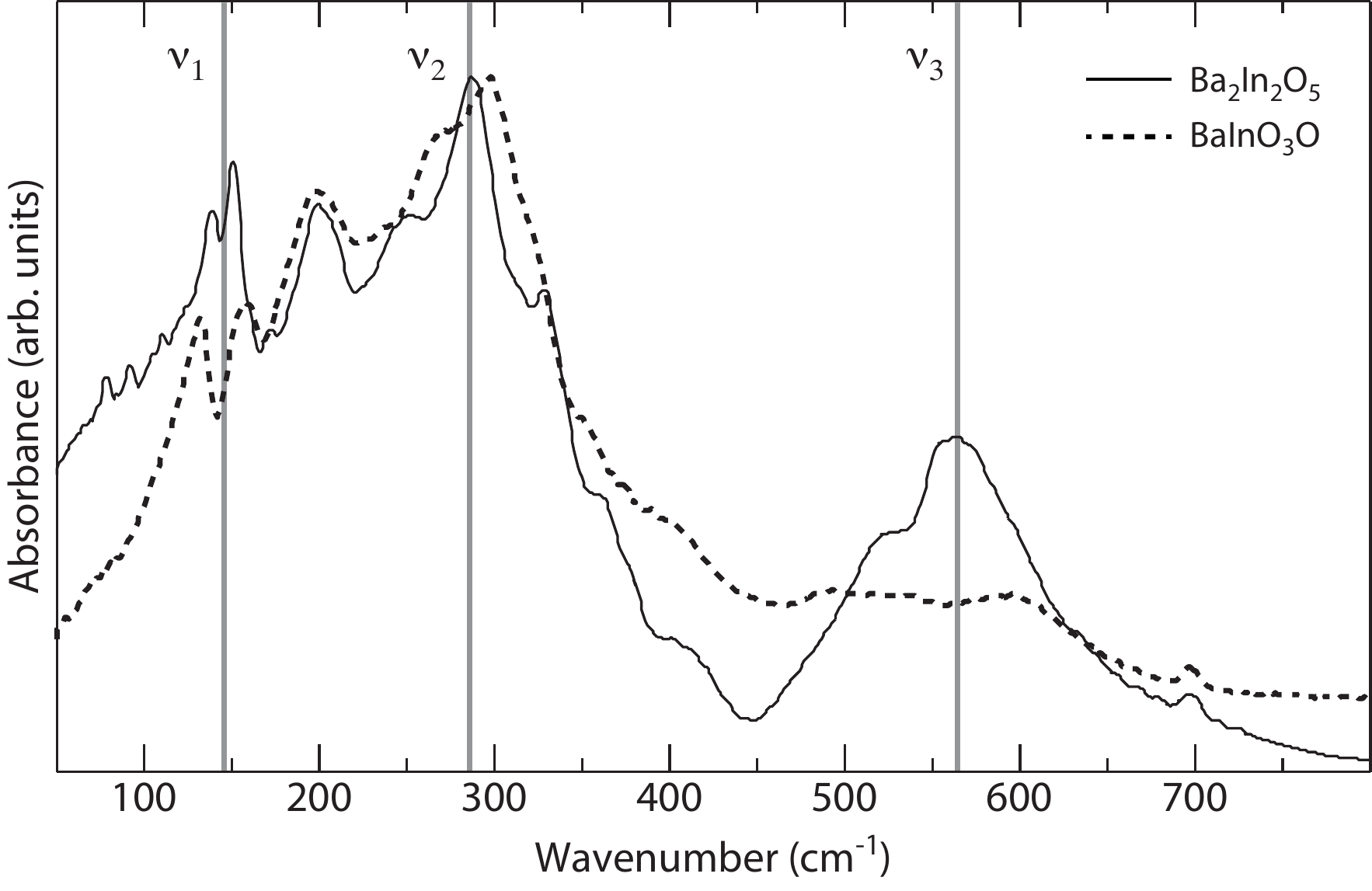}
\caption{IR spectra of \BIO (solid line) and \BIOH (dotted line).
Vertical lines indicate bands related to Ba-InO$_6$/InO$_4$ stretches ($\nu_1$), In-O-In bends ($\nu_2$), and In-O stretches ($\nu_3$), respectively. }
\label{fig:IR}
\end{figure}

The IR spectra of \BIO and \BIOH are shown in Fig.~\ref{fig:IR}.
The vertical lines indicate three main regions of bands, centered at around 140, 290, and 500 \cmi, respectively, which are assigned to Ba-InO$_6$/InO$_4$ stretches ($\nu_1$), O-In-O bends  ($\nu_2$), and In-O stretches ($\nu_3$), respectively.\cite{KAR08_Shortrange} 
The presence of several side bands (shoulders) on these main bands is a result of the non-cubic crystal symmetry.
In particular, we observe a strong band at around 200 \cmi, which has been assigned to a torsional mode of the oxygen lattice,\cite{KAR08_Shortrange}  as well as weaker features below 100 \cmi. 
Moreover, we observe that the spectrum of the hydrated material contains a smaller number of side bands than the spectrum of the dehydrated material and that the two lowest energy bands below 100 \cmi are not discernible, indicating a higher symmetry of the structure of the hydrated material.

\subsection{Raman spectra}

\begin{figure}[t]
\includegraphics[width=1\columnwidth]{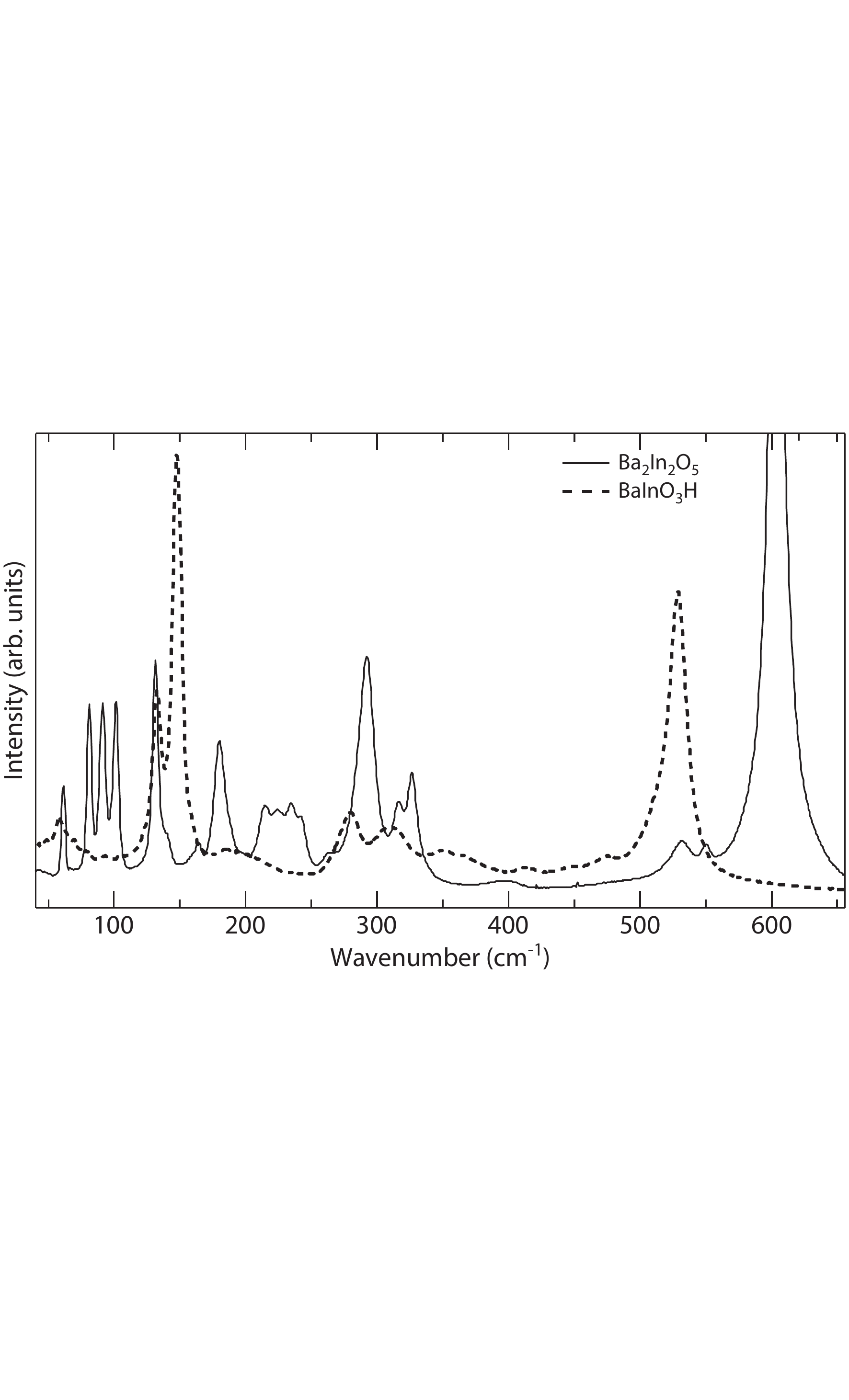}
\caption{Raman spectra of \BIO (solid line) and \BIOH (dotted line). 
}
\label{fig:Raman}
\end{figure}

The Raman spectra of \BIO and \BIOH are shown in Fig.~\ref{fig:Raman}. The spectrum of \BIO contains 23 distinguishable bands, which should be compared with the theoretical number of modes in the different brownmillerite structures: 51 in $Ibm2$, 48 in $Pnma$, and 30 in $Icmm$, see Table~\ref{tab:Modes}. 
Out of these, the strongest are five sharp bands below 140 \cmi, one strong band at 180 \cmi, a broad band composed of four components between 200 and 260 \cmi, a triplet of bands at around 300 \cmi, and a particularly pronounced band at 600 \cmi together with two weaker bands at its low-frequency side.
A comparison with the corresponding IR spectral assignment suggests that the bands below 140 \cmi are dominated by vibrations of the heavy Ba ions, the bands above 450 \cmi are related to In-O stretch modes, whereas the bands between 120 and 400 \cmi may be related to different tilt/bend motions of the InO$_{4}$ and InO$_{6}$ units present in the structure.\cite{Lazic}
 \begin{table*}
 {\footnotesize
\caption{The number of group theoretically allowed vibrational modes for \BIO and \BIOH. To better compare the two, we have separated the hydrogen derived modes in \BIOH.}
\begin{tabular}{l r r}
\hline
 & Raman & IR\\
\hline
\BIO (\emph{Ibm2}) & $13A_1+12A_2+12B_1+14B_2=51$ & $13A_1+12B_1+14B_2=39$\\
\BIO (\emph{Pnma}) & $13A_g+11B_{1g}+13B_{2g}+11B_{3g}=48$ & $15B_{1u}+13B_{2u}+15B_{3u}=43$\\
\BIO (\emph{Icmm}) & $9A_g+6B_{1g}+7B_{2g}+8B_{3g}=30$ & $11B_{1u}+8B_{2u}+8B_{3u}=27$ \\
\BIOH (\emph{P4/mbm}) - excl. H-modes & $4A_{1g}+2B_{1g}+3B_{2g}+6E_{g}=15$ & $6A_{2u}+13E_{u}=19$\\
\BIOH (\emph{P4/mbm}) - only H-modes & $3A_{1g}+3B_{1g}+3B_{2g}+6E_{g}=15$ & $3A_{2u}+6E_{u}=9$\\
\hline

\end{tabular}
\label{tab:Modes}
}
\end{table*}

Upon hydration, the Raman spectrum changes drastically, now with only three strong bands, located at around 135, 150, and 530 \cmi, respectively, as well as a broader region of overlapping bands in the range between 250 and 400 \cmi, present. In the same way as for the IR spectra, the overall reduction in the intensity of the Raman active modes signifies a more cubic structure in the hydrated phase, where we note that the proposed $P4/mbm$ symmetry contains 15 Raman active modes (Tab. \ref{tab:Modes}).
Following the assignment of the spectrum of the dehydrated material, the sharp 135 and 150 \cmi bands are most likely related to vibrations involving mainly Ba and/or In motions, the broad band between 250 and 400 \cmi to oxygen tilt/bend modes, and the strong 530 \cmi band to In-O vibrations.

\section{DFT calculations}

\subsection{Calculated INS spectra}

\begin{figure}
\includegraphics[width=1\columnwidth]{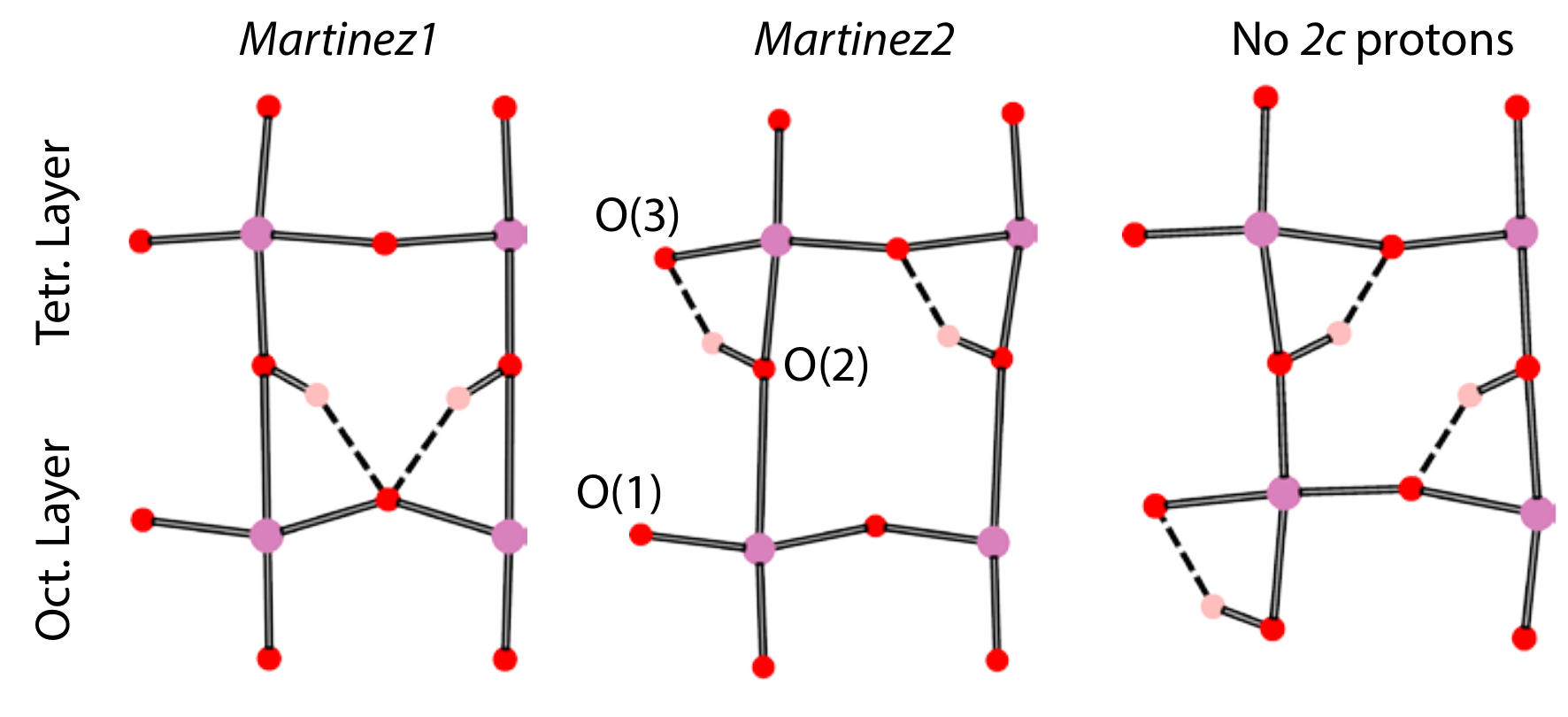}
\caption{H(1) proton positions on the $32y$ site in the different configurations investigated by DFT calculations. The illustrations are drawn within the plane that contains the H(1)-O hydrogen bonds in Fig. \ref{fig:Structure}. The octahedral (Oct.) and tetrahedral (Tetr.) layers according to Fig. \ref{fig:Structure} are also indicated.
}
\label{fig:Model}
\end{figure}

The INS spectrum contains, through the frequencies of the O-H wag modes, sensitive structural information connected to the proton positions in \BIOH. In order to find the proton configuration most compatible with the experimental INS data, DFT calculations were performed for three different low energetic proton configurations according to Martinez \emph{et al.}\cite{Martinez}, who considered 20 different structures characterized by different configurations of the protons in both the octahedral and tetrahedral layers.
The three proton configurations investigated here were chosen in order to discriminate between the vibrational signatures of hydrogen bonds from the O(2)-H(1) hydroxyl towards oxygens in either the octahedral layer, tetrahedral layer, or a combination thereof, see Fig. \ref{fig:Model}. 

The lowest energy structure (\emph{Martinez1}) is that shown on the left-hand side of Fig. \ref{fig:Structure} and in Fig. \ref{fig:Model} where lattice and geometry optimization resulted in a $P$1 structure with $a$ = 5.635, $b$ = 5.621, $c$ = 8.810 \AA, $\alpha$ = 90.078, $\beta$ = 90.174, $\gamma$ = 91.928 $^\circ$, V = 278.9 \AA$^{3}$ (\emph{cf.} the initial structure $a$ = $b$ = 5.915, $c$ = 8.999 \AA, $\alpha  =  \beta =  \gamma$ = 90 $^\circ$, V = 314.8 \AA $^{3}$). The marked change in volume is caused by the In$-$O$-$In bonds becoming bent instead of linear. The H(2) proton in the 2$c$ site has moved to the 4$h$ site, $i.e.$ towards an oxygen to give a reasonable O$-$H bondlength of 0.984 \AA\  and the H(1) proton at the 16$l$ site has relaxed to a new position denoted as 32$y$â with an O$-$H bondlength of 0.973 \AA.

\begin{figure}
\includegraphics[width=1\columnwidth]{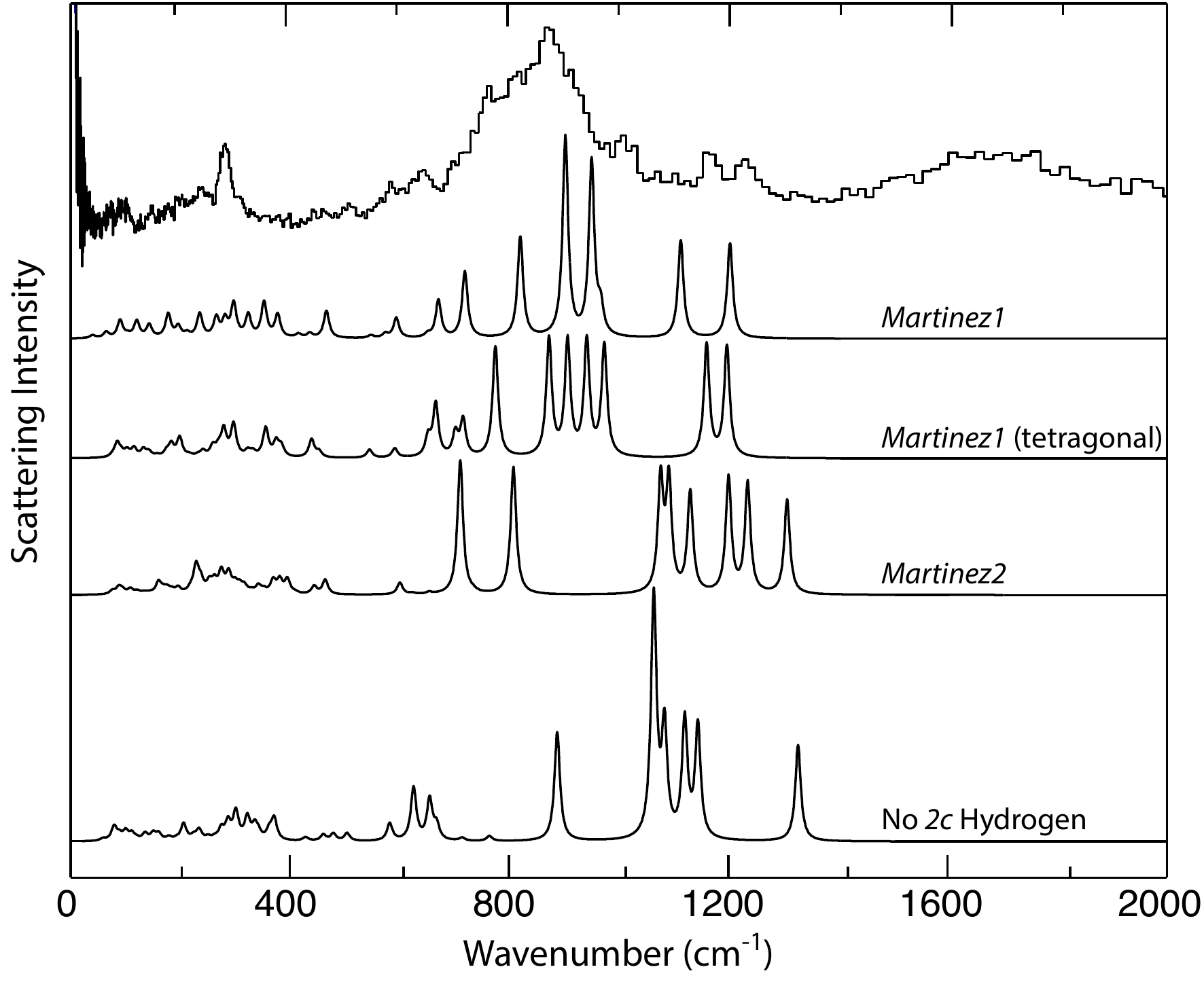}
\caption{Comparison of the INS spectrum of \BIOH with the first order spectrum calculated for the structures in Fig. \ref{fig:Model}.
The calculated vibrational INS frequencies are shown convoluted with a 12 \cmi FWHM Lorentzian line shape.
}
\label{fig:INSCalc}
\end{figure}

A comparison between the experimental INS spectrum of \BIOH and that calculated based on the $Martinez1$ structure is shown in Fig. \ref{fig:INSCalc} 
and suggests that the broad O-H wag band at around 800 \cmi corresponds to several O-H wag modes, originating primarily from $4h$ proton motions, whereas the high-frequency shoulder at 960 \cmi and the two well defined bands at around 1200 \cmi involve only the $32y$ protons.
A free geometry optimization in the \emph{Martinez1} structure reduces the $P4/mbm$ symmetry to $P$1, but experimentally a tetragonal cell is found. Starting from the same initial structure but requiring the cell lattice (but not its contents) to be tetragonal, resulted in a very similar structure that was 0.056 eV higher in energy with cell parameters $a = b =$ 5.622, c = 8.825 \AA, $\alpha  =  \beta =  \gamma$ = 90 $^\circ$, V = 278.9 \AA $^{3}$. 
The corresponding calculated INS spectrum, denoted $Martinez1 (tetragonal)$, is included in Fig. \ref{fig:INSCalc} and as can be seen it is similar to what was obtained in the freely optimized $Martinez1$ structure with only small frequency shifts between the two. 

To further investigate the possible proton positions, the next lowest energy configuration ($Martinez2$) found by Martinez  \emph{et al.}\cite{Martinez} was also investigated. 
Geometry optimization produced the structure shown in the center of Fig.~\ref{fig:Model} and this structure was found to be slightly lower in energy by 0.048 eV than the $Martinez1$ structure. 
The most likely reason for this is that the hydrogen bonds are much stronger in this case ($\sim$1.7~\AA) compared to the $Martinez1$ structure ($\sim$1.9~\AA), and as a consequence the O-H wag modes are now found at significantly higher energies than in the experimental spectrum, see Fig.~\ref{fig:INSCalc}.
This suggests that the $Martinez2$ structure is not a correct description of the structure of the real material.

Finally, we investigated a structural arrangement characterized by three 32$y$ sites and one 4$h$ site occupied and this arrangement is shown on the right-hand side of Fig.~\ref{fig:Model}.
This structural arrangement was however found to be significantly higher in energy (0.4 eV) than both the $Martinez1$ or $Martinez2$ structures, pointing towards the need to occupy the 2$c$ site. 

To conclude, a satisfactory agreement between experiment and calculated INS spectra is obtained only for the $Martinez1$ structure, as even relatively small displacements of the positions of protons away from the $Martinez1$ structure was found to result in spectra quite incompatible with the experiment.
The main source of discrepancy between the experimental and calculated spectra can most likely be attributed to broadening effects induced by local fluctuations in the short-range structure around the protons in the material, although some part of the band broadening is related also to the vibrational lifetimes.

\subsection{Calculated IR spectra}

\begin{figure}
\includegraphics[width=1\columnwidth]{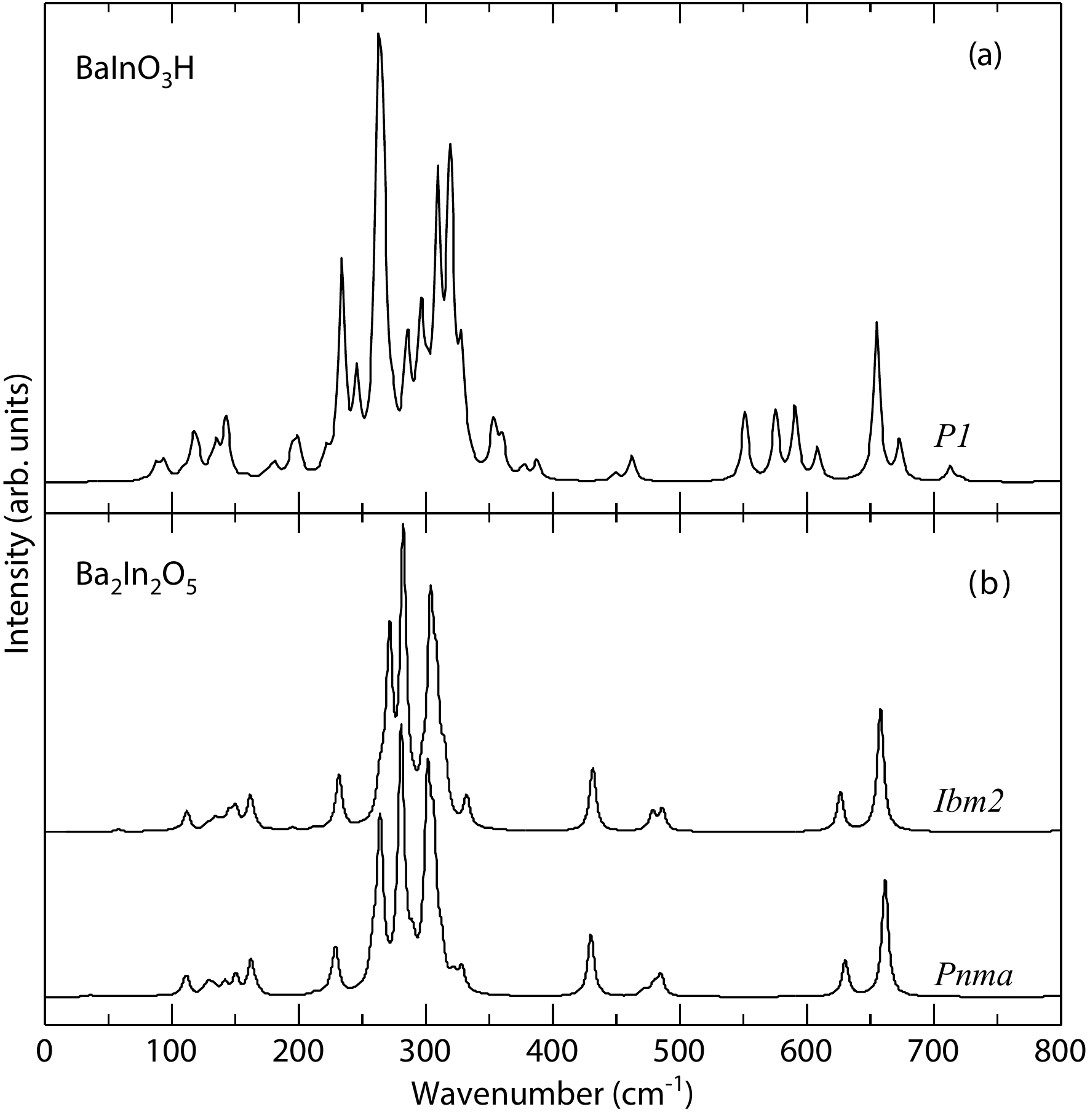}
\caption{Calculated IR spectra of (a) \BIOH and (b) \BIO, using the space groups in the calculations as indicated.
The spectra have been convoluted with a 3 \cmi FWHM Lorentzian. 
}
\label{fig:DFT_IR}
\end{figure}

The calculated IR spectra of \BIOH and \BIO, are shown in Fig.~\ref{fig:DFT_IR}. 
The spectra have been obtained from geometry optimized structures with the space group symmetries $Ibm2$ and $Pnma$ (\BIO) and $P1$ with the $Martinez1$ proton configuration (\BIOH). 

Considering first the spectra of the dehydrated material [Fig.~\ref{fig:DFT_IR}(b)], we observe that the two space group symmetries yield spectra overall very similar to each other.
From the displacement vectors, extracted from the calculations, we find that the spectra may be divided into three well separated regions of bands, related to Ba-InO$_6$/InO$_4$ vibrations (0--200 \cmi), In-O-In bend modes (200--400 \cmi), and In-O stretch modes (600--800 \cmi), respectively.
These findings support our empirical assignments of the $\nu_1$, $\nu_2$ and $\nu_3$ bands in the experimental spectra as discussed above. 
Surrounded by the bend (200--400 \cmi) and stretch (600--800 \cmi) regions, we also observe a distinct band at approximately 430 \cmi, related to In-O-In bend vibrations, and a triplet of bands at around 490 \cmi, related to two weaker bend modes and one stronger stretch mode.
For a complete list of the calculated atomic motions and frequencies, as well as a corresponding phonon assignment of the experimental spectra of \BIO, which will be discussed more below, see Table~\ref{tab:assignment}.

Considering next the spectrum of the hydrated material [Fig.~\ref{fig:DFT_IR}(a)], this is overall similar to the spectra of the dehydrated material, with the same spectral regions of bands clearly distinguishable in all three spectra. 
This suggests that the spectral changes that occur upon hydration are in large caused by selection rules and vibrational cross sections rather than in shifts of vibrational frequencies. 
The similarity between the calculations are consistent with the experimental observation of rather small differences in the IR spectra of \BIO and \BIOH as can be seen in Fig. \ref{fig:IR}.

\label{sec:DFT}

\begin{table*}[t]
{\footnotesize
\caption{Phonon assignments for \BIO, based on the $Pnma$ crystal symmetry. Primed and un-primed numbers denote IR and Raman active phonons, respectively.}
\begin{tabular}{p{1.25cm} p{1.75cm} p{1.75cm} p{8cm} p{1.65cm}}
\hline
Mode & Exp. freq. (\cmi) & Calc. freq. (\cmi) & Atomic motion & Symmetry of motion\\
\hline
1&61&55& In(2)O$_4$ rotation around c-axis & $A_g$\\
1'&79&-& -& - \\
2&82&77& Ba-In(2)O$_4$ vibration within ab-plane&$A_g$\\
2'&92&93& Ba vibration within ab-plane&$B_{1u}$\\
3&92&85& Ba vibration within ab-plane&$B_{1g}$ \\
3'&111&113& In(2)O$_4$-Ba vibrations within ab-plane&$B_{3u}$ \\
4&102&95& Ba vibration within ab-plane&$B_{3g}$ \\
4'&139&152& In(1)O$_6$-Ba vibrations within ab-plane&$B_{1u}$ \\
5&132&138& Ba-In(2)O$_4$ vibration &$A_g$ \\
5'&151&164& In(1) vibrating against In(2)-O(3) layer&$B_{2u}$ \\
6&140&150& Rotation of In(2)O$_4$ around the O(3) axis &$B_{3g}$\\ 
6'&172&-& - &-\\
7&160&-& -\\
7'&200&230& In(2)O$_4$ bending& $B_{3u}$\\
8&164&-& -\\
8'&250&265& In(2)O$_4$ bending & $B_{3u}$\\
9&180&173& Ba-In(2)O$_4$ vibration along c& $A_g$\\
9'&287&282& In(2)O$_6$ bending& $B_{1u}$\\
10&187&-& -& -\\
10'&330&302,307&In(2)O$_4$ + In(1)O$_6$ out of plane/in-plane, in-phase, bend& $B_{2u}$,$B_{3u}$\\
11&197&-& -& -\\
11'&365&430& In(2)-O(3)-In(2)+In(1)-O(1)-In(1) out of plane, out of phase, bend& $B_{2u}$\\
12&215&230& In(1)-O(2)-In(2) bending & $A_g$, $B_{1g}$\\
12'&415&480,485& Asym. In(1)-O(1) stretch &$B_{2u}$,$B_{1u}$\\
13--15&224--243&235--252& In(2)O$_4$ and In(1)O$_6$ tilts & $A_g$, $B_{1g}$\\
13'&525&630&Out of phase asym. In(1)-O(2) and sym. In(2)-O(2) stretch & $B_{1u}$\\
14'&563&662&In phase asym. In(1)-O(2) and In(2)-O(2) stretch & $B_{2u}$\\
16&266&267& Sym. In(2)O$_4$ bend& $A_g$\\
17&292&293& In(1)O$_6$ tilt& $A_g$\\
(18,19)&(317,327)& 350.5& O(1)-In(1)-O(2) scissor & $A_g$\\
20&398&443& In(1)-O(1)-In(1) bend& $B_{3g}$\\
(21,22)&(532,550)&542& In(1)-O(1) stretch& $B_{3g}$\\
23&603&679& In(2)-O(2) stretch& $A_g$\\
\hline
\end{tabular}
\label{tab:assignment}
}
\end{table*}

\subsection{Calculated Raman spectra}

\begin{figure}
\includegraphics[width=1\columnwidth]{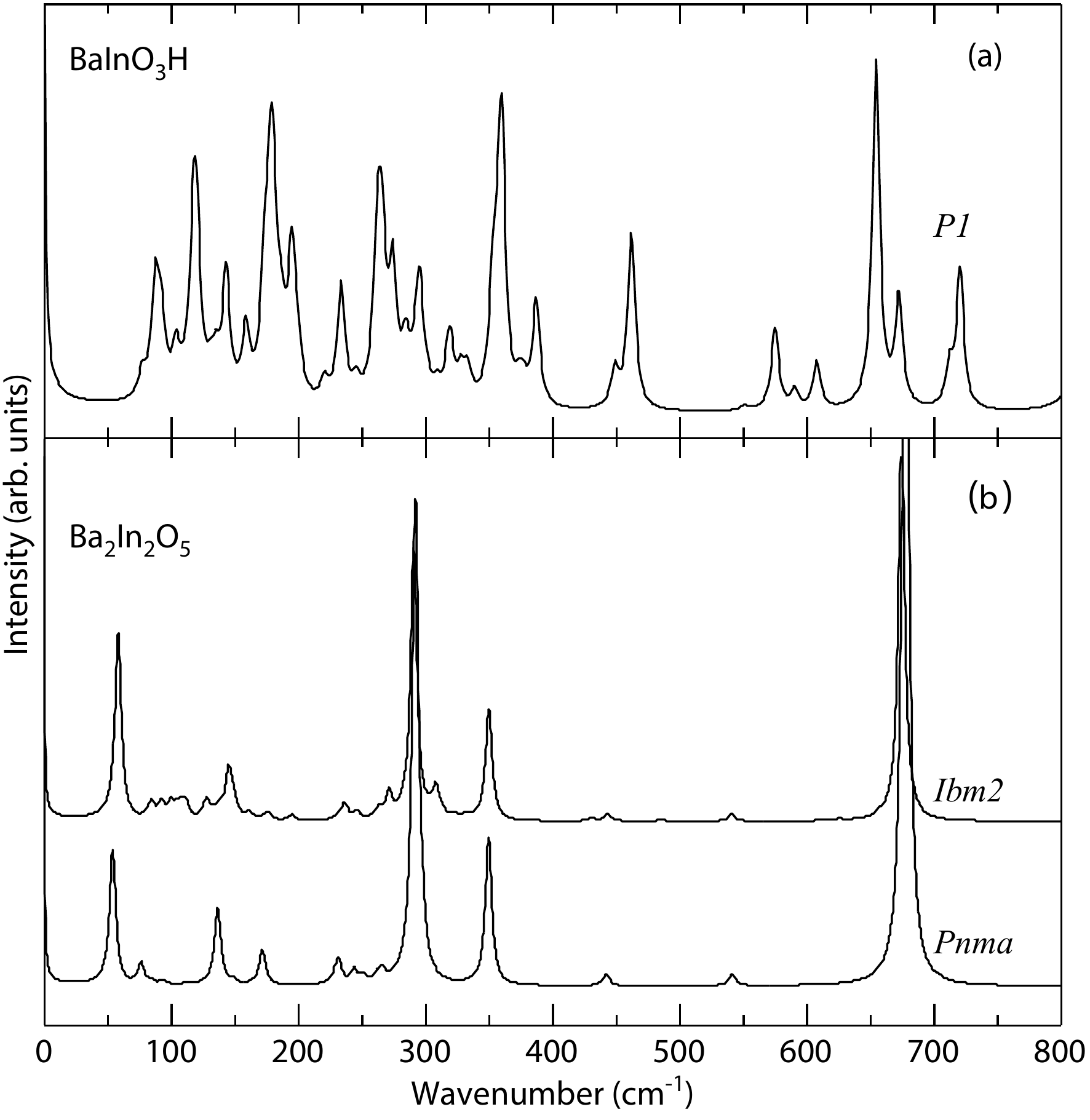}
\caption{
Calculated Raman spectra of (a) \BIOH and (b) \BIO. 
The spectra have been convoluted with a 3 \cmi FWHM Lorentzian. 
}
\label{fig:DFT}
\end{figure}

Figure~\ref{fig:DFT} shows the calculated Raman spectra of \BIOH and \BIO, both using the same symmetries as in the IR case described above.
For the dehydrated material [Fig.~\ref{fig:DFT}(b)], the Raman spectra are overall similar for the two space group symmetries, in agreement with the IR results. 
The Raman spectra are dominated by Ba modes below 180~\cmi, followed by InO$_4$ and InO$_6$ tilt modes between 180 and 300 \cmi, In-O-In bend modes in the range between 300 and 450 \cmi, and In-O stretch modes above 500 \cmi, \textit{cf.} Tab.~\ref{tab:assignment}.
In addition, the Raman spectrum for $Ibm2$ includes two bands between 130 and 180 \cmi, which are assigned to In(1) related vibrations. 
According to group theory, the In(1) site should not contribute to any modes in the centrosymmetric $Pnma$ structure, and indeed this is also the case.
The lowest-frequency band at approximately 55 \cmi is assigned to In(2)O$_4$ tilt vibrations. 

Upon hydration, the Raman spectrum changes considerably [see Fig. \ref{fig:DFT}(a)], now containing a significantly larger number of bands
A large effect on hydration was also observed in the experimental Raman spectra [Fig.~\ref{fig:Raman}], however here the number of modes was significantly decreased, pointing towards an increase in the symmetry of the crystal structure. 
The discrepancy between theory and experiment can be understood from the fact that the lack of symmetry in the $P1$ space group - as used in the calculations - makes all 69 optical modes Raman active, and realizing that the $P1$ symmetry is an artifact from the particular proton arrangement needed in order to avoid fractional occupancies in the calculations.

\section{Discussion}

\begin{figure}[t]
\includegraphics[width=1\linewidth]{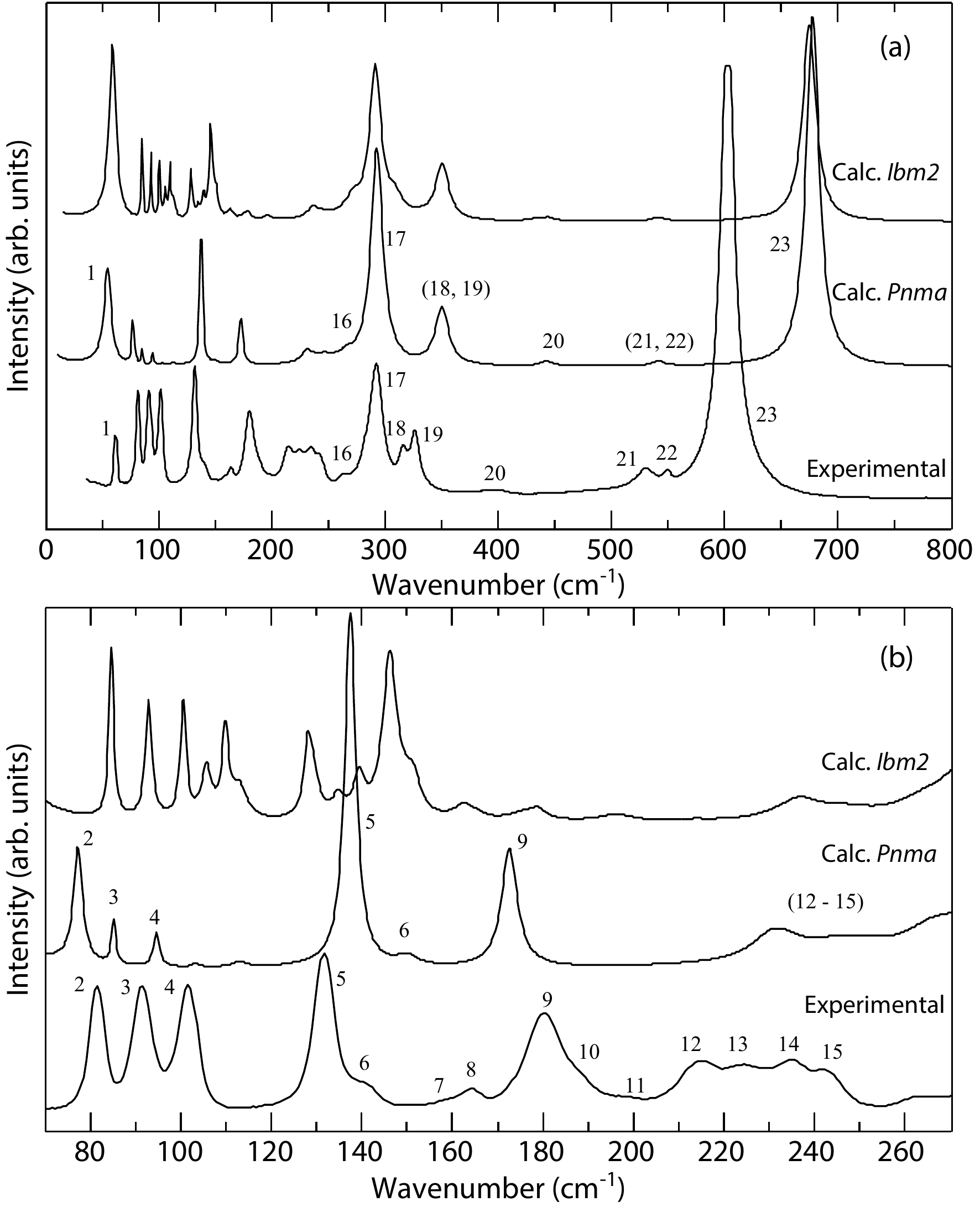}
\caption{Comparison between the calculated and measured Raman spectra of \BIO in the spectral range (a) 0--800 \cmi and (b) 75--270 \cmi. 
The best agreement between calculated and experimental spectra is found for the $Pnma$ structure and therefore it is according to this space group symmetry the phonon assignment has been made. 
The numbers refer to the phonon assignment presented in Tab.~\ref{tab:assignment}.
}
\label{fig:DryModes}
\end{figure}

By combining the results obtained from the INS,  IR and Raman spectra and DFT calculations we can distinguish several features related to the short-range structures of \BIO and its hydrated, proton conducting form \BIOH.
In particular, we find that the calculated Raman spectra for \BIO, especially in the 50--200 \cmi range, are very sensitive to the structural model used and vary significantly between the investigated space group symmetries, $Ibm2$ and $Pnma$, and may therefore be exploited as a means to elucidate the true short-range structures present.
These points are illustrated in Fig.~\ref{fig:DryModes}, where the experimental Raman spectrum is shown together with the calculated $Pnma$ and $Ibm2$ spectra, which now have been homogeneously broadened to more closely resemble the experimental spectrum. 

\begin{figure}[t]
\includegraphics[width=1\linewidth]{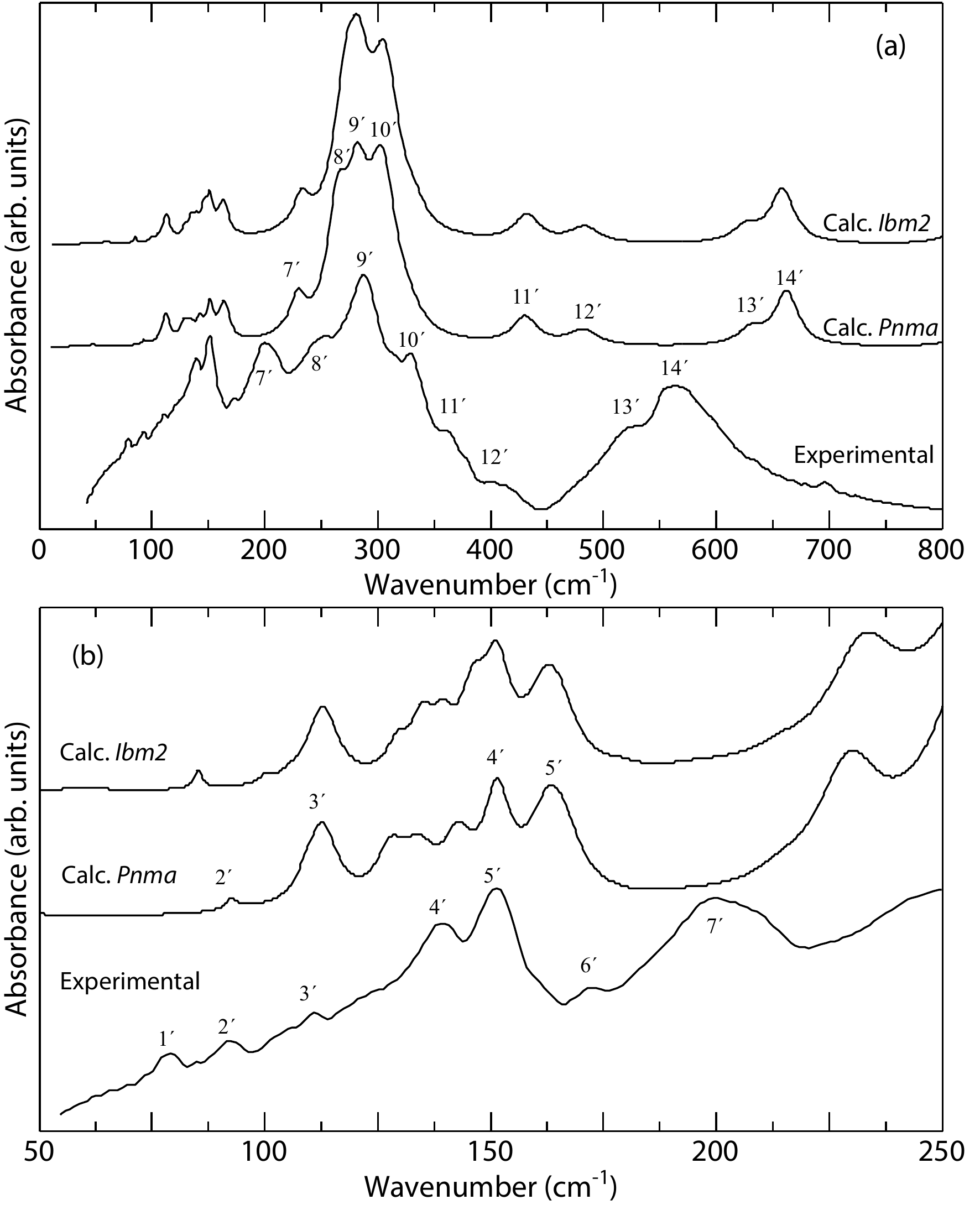}
\caption{Comparison between the calculated and measured IR spectra of \BIO in the spectral range (a) 0--800 \cmi and (b) 50--250 \cmi. 
The numbers refer to the phonon assignment presented in Tab.~\ref{tab:assignment}.
}
\label{fig:DryModes-IR}
\end{figure}

Considering first the 70--120 \cmi range, a key result is that the calculated Raman spectra for $Pnma$ is in agreement with the experimental spectrum (in a presumed $Icmm$ structure), whereas the calculated spectrum for $Ibm2$ symmetry contains at least five bands in this range, suggesting that the $Ibm2$ symmetry is not a correct description of the structure of \BIO. 
At slightly higher frequencies (120--200 \cmi), the $Ibm2$ calculation gives a plethora of phonon modes, activated by the non-centrosymmetric In position, whereas for $Pnma$ only three bands, labeled 5, 6, 9, are observed. 
These three bands are also the three strongest bands observed experimentally, although one should note that the measured spectrum also contains additional, weaker, modes, labeled 7, 8, 10, and 11, that are absent in the calculated spectra for $Pnma$ symmetry, as well as significantly stronger Raman activity in the range between 200 and 250 \cmi. 
$Ibm2$ can thus be excluded as being the space-group symmetry in \BIO. As $Pnma$ is similarly excluded on the basis of diffraction\cite{Speakman}, we interpret these results as a consequence of $Icmm$ symmetry on the local scale. 
This interpretation is further supported by the differences in Raman intensity of  the low-frequency InO$_4$ band at 61 (exp.) and 55 (calc.) \cmi, related to an In(2)O$_4$ in-plane tilt mode, whose reduced intensity as compared to both calculations is interpreted as due to the decreased coherence between adjacent InO$_4$ layers in $Icmm$. 	
It follows that the $Icmm$ symmetry, as proposed on the basis of neutron diffraction, is not due to uncoupled $Ibm2$ domains on the local scale but rather the result of true randomness of the orientation of successive tetrahedral layers.

In Fig.~\ref{fig:DryModes-IR} is shown an analogous comparison between the experimental and calculated IR spectra, where a larger broadening than for the computed Raman spectra is necessary to mimic the experimental data. 
One should note that due to the very broad features in the experimental IR spectrum it is difficult to accurately relate the calculated modes to individual peaks, and in particular the assignment of peaks below 230 \cmi should be regarded with some caution.
Together with the fact that the selection rules for IR spectroscopy result in $Ibm2$ and $Pnma$ spectra that are very similar to each other, we conclude that Raman spectroscopy is a more suitable technique for distinguishing between possible short-range structures of brownmillerites. 

Considering now the hydrated material, \BIOH, we note that the average $P4/mbm$ structure as deduced from neutron diffraction, is a highly symmetric structure with a low number of Raman and IR active modes (see Tab. \ref{tab:Modes}) consistent with the experimental observations shown in Fig.~\ref{fig:IR} and Fig.~\ref{fig:Raman}. 
For this reason, the symmetry reduction to $P1$ in the modeled \BIOH unit cell necessarily implies a weak agreement between the experimental and modeling results.
In the case of INS, however, which is not subject to symmetry induced selection rules as in Raman and IR spectroscopy, we can expect much more accurate results from the calculations. 
The experimental INS spectrum of \BIOH indeed suggests a proton arrangement corresponding to the $Martinez1$ structure, with both the main band at 860 \cmi and the weaker doublet around 1200 \cmi present in the calculations. 
The calculations further show that the hydrogen bonding within the cell is relatively weak with O-H$\cdots$O distances mostly in the range 1.9--2.0 \AA, as shorter distances result in many of the modes located at higher wavenumbers than is experimentally observed. 
The three sharp modes at 1008, 1163 and 1227 \cmi involve only $4h$ protons, while the broad band peaked at 860 \cmi contains four sub-bands, out of which three only involves $32y$ protons. 

Turning again to the Raman spectrum of \BIOH, we notice that a distinct feature of the $Martinez1$ structure is a long-range distortion of the hydrogen bonding O(1) oxygens caused by the ordered H(1) proton arrangement. This, in turn, induces a long-range non-centrosymmetric distortion of the In(1)-O$_6$ octahedra, which should be observable in the Raman spectrum by the activation of previously Raman inactive modes related to the In(1) Wyckoff position. 
From the computational results, these In(1) related modes are found to lie in the spectral range between 140 and 160 \cmi. 
As the only spectral feature in \BIOH that lacks a counterpart in the dry \BIO sample is the intense 150 \cmi vibration, we assign this peak to such an In(1) related mode caused by the long-range coherence of the $Martinez1$ proton arrangement.
 
As a final remark we note that the difference in hydrogen-bond strength between the H(1) protons in the $32y$ position (weak hydrogen bonding) and H(2) protons in the $4h$ position (strong hydrogen bonding) points towards a difference in proton mobility in these two sites.
This is because the higher the degree of hydrogen bonding the higher the rate of proton transfer, which is a hydrogen-bond mediated process, whilst the -OH reorientational motion requires the breaking of such bonds.
This would suggest that the rate of proton transfer is higher for H(2) than for H(1), and vice versa for the reorientational rate. 
On the other hand, the full occupation of H(2) protons on the $4h$ position should hinder significantly the diffusion of protons within the In(2)--O(3) plane containing the nearest oxygen neighbors to which the H(2) protons form strong hydrogen bonds.
This would rather suggest that the proton transfer rate is higher for the more weakly hydrogen bonded H(1) protons. 
However, upon dehydration with increasing temperature it might then be that the rate of H(2) proton transfer events increases at a rate that is a function of the H(2) occupancy. 
This opens exciting new questions such as: does the full occupation on the $4h$ position mean that the proton conductivity is governed by the $32y$ site, and if so, is there any optimum occupancy?
Further research along these lines, in particular by quasielastic neutron scattering, which can give information about the mechanistic detail of local diffusional proton motions, may be very rewarding.

\section{Summary and Conclusions}

In summary, we have investigated the vibrational spectra and short-range structure of the brownmillerite-type oxide \BIO and its hydrated form, using INS, Raman and IR spectroscopies together with DFT calculations.
Previous results obtained from neutron diffraction measurements on \BIO suggest a structure of $Icmm$ symmetry, characterized by random handedness of successive layers of InO$_{4}$ tetrahedra, however it has not been clear to which extent such a randomness is due to uncoupled $Ibm2$ domains or actual random handedness on the local scale.
In the present work, we show that the $Ibm2$ symmetry can be ruled out. 
Instead, we find a close agreement with the $Pnma$ symmetry, although we note that the experimental and calculated spectra are not exactly the same, suggesting that $Icmm$ is the true structure of \BIO also on a more local length-scale than that previously probed with neutron diffraction. 
The main source of deviations between the experimental and calculated spectra relate to the low-frequency part below 250 \cmi, which is significantly different for the $Ibm2$, $Pnma$, and $Icmm$ symmetries and may therefore be exploited as a means to distinguish between the handedness of different brownmillerite type structures in general.

For the hydrated, proton conducting, material \BIOH, a comparison of the experimental and calculated INS spectra based on fully localized (no partial occupancies) structures of $P1$ symmetry suggests the protons to mainly order according to the $Martinez1$ structure. Consistent with this we observe a sharp Raman active peak at 150 \cmi that we attribute to the non-centrosymmetric In(1) position caused by the $Martinez1$ proton arrangement. 
Finally, our results indicate that the mechanism of proton conduction may depend strongly on the level of hydration of the material.

\acknowledgements
We thank the group of Sten Eriksson at Chalmers University of Technology, for the access to sample preparation facilities.
This research was primarily funded by the Swedish Research Council. 
The STFC Rutherford Appleton Laboratory is thanked for access to neutron beam facilities.

\providecommand*\mcitethebibliography{\thebibliography}
\csname @ifundefined\endcsname{endmcitethebibliography}
  {\let\endmcitethebibliography\endthebibliography}{}

\begin{figure*}
\includegraphics[width=0.6\linewidth]{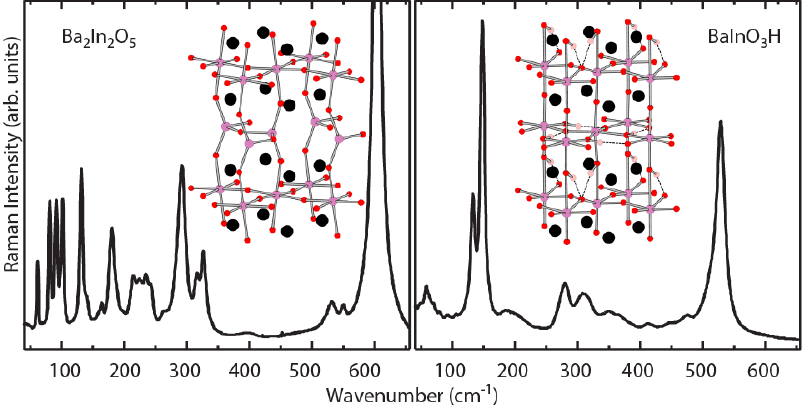}
\caption{\textbf{TOC Summary and Graphic:} \\ Raman spectra reveal short-range structure and local coordination of protons in brown- \\ -millerite-type oxide \BIO and its hydrated proton-conducting analogue \BIOH.}
\label{TOC}
\end{figure*}

\end{document}